\documentclass[12pt]{article}
\usepackage{graphicx,latexsym}
\usepackage{amsmath}
\usepackage{amssymb}
\usepackage{float}
\usepackage[utf8]{inputenc}
\usepackage{color}
\usepackage{cite}
\usepackage{authblk}

\begin{document}

\title{Numerical treatment of nonlinear Fourier and Maxwell-Cattaneo-Vernotte heat transport equations}

\author{R. Kovács$^{123}$, P. Rogolino$^4$}

\affil{
$^1$Department of Energy Engineering, Faculty of Mechanical Engineering, BME, Budapest, Hungary \\
$^2$Department of Theoretical Physics, Wigner Research Centre for Physics,
Institute for Particle and Nuclear Physics, Budapest, Hungary \\
$^3$Montavid Thermodynamic Research Group \\
$^4$Department of Mathematics and Computer Sciences, Physical Sciences and Earth Sciences, University of Messina, Messina, Italy}



\date{\today}
\maketitle

\begin{abstract}
The second law of thermodynamics is a useful and universal tool to derive the generalizations of the Fourier's law. In many cases, only linear relations are considered between the thermodynamic fluxes and forces, i.e., the conduction coefficients are independent of the temperature. In the present paper, we investigate a particular nonlinearity in which the thermal conductivity depends on the temperature linearly. Also, that assumption is extended to the relaxation time, which appears in the hyperbolic generalization of Fourier's law, namely the Maxwell-Cattaneo-Vernotte (MCV) equation. 
Although such nonlinearity in the Fourier heat equation is well-known in the literature, its extension onto the MCV equation is rarely applied. Since these nonlinearities have significance from an experimental point of view, an efficient way is needed to solve the system of partial differential equations. In the following, we present a numerical method that is first developed for linear generalized heat equations. The related stability conditions are also discussed. 
\end{abstract}


\section{Introduction}

The generalization of classical constitutive equations -- such as the Fourier, Hooke, or Newton equations -- is needed from both theoretical and practical points of view. There are many experimental evidence on the existence of phenomena beyond these classical models \cite{Tisza38, Pesh44, JacWalMcN70, Gyarmati70b, FulEta14m1, SzucsFul18a, SzucsFul18b, CarMorr72a, LebCloo89, SellEtal16b, Verhas97, Str11b}. Due to the complexity of the description of non-classical phenomena, we focus only on the heat conduction equations for rigid media.
Most of the heat conduction problems are described and studied using Fourier's law 
 which relates linearly the temperature gradient $\nabla T$
to the heat flux $\mathbf q$:
\begin{equation}\label{fourier}
\mathbf q=-\lambda \nabla T
\end{equation}
where $\lambda$ represents the thermal conductivity, which, in general, depends on the temperature. Here, $\nabla$ is the nabla operator.
It is well known that this equation, at least with constant $\lambda$, implies infinite speed of propagation.
Consequently, the constitutive  \eqref{fourier} leads to a parabolic equation for the temperature. Thus it 
fails to describe the heat transfer mechanism over short time scales with a large temperature gradient. Such a situation occurs under different circumstances, for instance, in laser heating processes.

In order to eliminate the paradox of infinite velocity, one can apply the so-called Maxwell-Cattaneo-Vernotte (MCV) constitutive equation \cite{Max1867, Cat1, Cattaneo58, Vernotte58, Gyar77a},
\begin{equation}
\tau \partial_t \mathbf q + \mathbf q = -\lambda \nabla T, \label{MCV}
\end{equation}
where the coefficient $\tau$ is the relaxation time. That was the first extension of Fourier's law \eqref{fourier} in which the time derivative term introduces an intertia - a time lag - that makes the model hyperbolic. 
When $\tau$ is negligibly small, then the heat flux $\mathbf q$ becomes proportional with the temperature gradient $\nabla T$, i.e., eq.~(\ref{MCV}) reduces to the Fourier's law. We draw attention to the non-equilibrium thermodynamical background of the MCV model, which requires the heat flux to be an independent state variable following the approach of Extended Irreversible Thermodynamics (EIT) \cite{JouVasLeb88ext, LebEta11a, LebEtal08b, JouEtal99, Lebon14}. This structure is also analyzed in \cite{CimKovVanRog, RogCim}, and more general constitutive equations are derived, reproducing both parabolic and hyperbolic equations. Consequently, the thermal conductivity and the relaxation time might depend on the heat flux, too. 
The MCV equation \eqref{MCV} proved to be useful in modeling low-temperature wave propagation, called `second sound' \cite{FairLaWill47, McN74t, Gyar77a, NarDyn72a, Naretal75, Cimmelli09nl}.
 It describes finite speed of propagation $v=\sqrt{\alpha/\tau}$, with $\alpha=\lambda/(\rho c_v)$ being the thermal diffusivity, $c_v$ is the isochoric specific heat and $\rho$ is the mass density.  
 
Beside the constitutive equation, it is needed to consider the balance of internal energy $e$ as well to obtain a complete system of partial differential equations,
\begin{equation}
\rho c_v\partial_t T + \nabla \cdot \mathbf q = 0, \label{inten}
\end{equation}
in which we consider no source terms and the $e=c_v T$ constitutive relation (state equation) for the internal energy is exploited. Furthermore, $\partial_t\{.\}$ denotes the partial time derivative and $\nabla\cdot\{.\}$ is the divergence operator. {We note that the source terms are neglected only for the sake of simplicity. Although they have importance in many situations, they do not contribute to our investigations now. It does not restrict the numerical solution method.}

From an experimental point of view, it is proved that the MCV equation is still not enough, e.g., it is inadequate for wave propagation in non-metallic crystals; and further extensions are required. One important example is the Guyer-Krumhansl equation \cite{GuyKru2},
\begin{equation}
\tau \partial_t \mathbf q + \mathbf q = -\lambda \nabla T + l^2 \Delta \mathbf q, \label{GK}
\end{equation}
in which a nonlocal term (the Laplacian of $\mathbf q$) appears and $l$ is a new
phenomenological coefficient.
Originally, Guyer and Krumhansl derived this model based on the linearization of the Boltzmann equation \cite{GuyKru66a1, GuyKru66a2}, and proved to be an outstanding model to predict the existence of second sound in solids. However, it can also be derived in the framework of non-equilibrium thermodynamics using internal variables \cite{KovVan15, BerVan17b}, and current multipliers \cite{Nyiri89, Nyiri91} that does not restrict its validity to low-temperature problems.  It is successfully applied in the evaluation of room temperature experiments as well \cite{Botetal16, Vanetal17}. Moreover, due to the nonlocal term, its numerical solution required a particular spatial discretization to embed the boundary conditions appropriately \cite{RietEtal18}.

In the following, we restrict ourselves to the Fourier and MCV equations. In most cases, the coefficients $\tau$ and $\lambda$ are constant and independent of the  temperature, which is acceptable in many situations. However, as some experimental evaluation shows, both parameters can depend on the temperature \cite{ColNew88nl, Jordan15, Romano}, especially in the low-temperature domain. Here, we consider  only
\begin{subequations}
\begin{eqnarray}\label{ltdepT}
&\lambda(T)=\lambda_0 + a(T-T_0),\\
&\tau(T)=\tau_0 + b(T-T_0) \label{ltdepT_a}
\end{eqnarray}
\end{subequations}
forms for the material parameters, which are the simplest and experimentally relevant dependencies, with $ \lambda_0$ and $\tau_0$ being the thermal conductivity and relaxation time at the reference (or initial) temperature $T_0$ respectively, and the parameters $a, b$ {could be both positive and negative. We emphasize that the resulting thermal conductivity and relaxation time must remain positive at the end; hence their values are limited in this sense.}

Usually, only the temperature is used as a primary field variable, that is, the heat flux is often eliminated from the system \eqref{fourier} (or \eqref{MCV}, accordingly) and  \eqref{inten}.This elimination introduces nonlinear terms, e.g., $\left(\frac{\partial T}{\partial x}\right)^2$ in the 1D form of Fourier heat equation:
\begin{equation}
\rho c_v \partial_t T = \partial_x (\lambda (T) \partial_x T).
\label{nonF}
\end{equation}
Despite that the Kirchhoff transformation is developed for such cases \cite{Kirc1894b, Grof18}, it is not advantageous for extended heat equations. Even in the simplest generalizations, the temperature-dependent material parameters would result in an unreasonably complicated form.  Furthermore, the implementation of practically useful boundary conditions, e.g., time-dependent heat flux, would be extremely difficult if even possible. Hence without eliminating the heat flux, we can avoid these nonlinear terms while preserving the physical meaning, and it is easier to prescribe the boundaries, too.  For numerical calculations, it is more convenient to use both variables, i.e., the temperature and the heat flux together. Indeed, this is the basis of the earlier developed numerical scheme that proposes a staggered spatial field for appropriate treatment of boundary conditions \cite{RietEtal18}. Thus we use the balance of internal energy (eq.~(\ref{inten})) together with the constitutive equation, eq.~(\ref{MCV}) with $\tau=0$ (Fourier) and $\tau \neq 0$ (MCV). As a consequence of \eqref{ltdepT}-\eqref{ltdepT_a}, the propagation speed $v=\sqrt{\lambda(T)/(\rho c_v \tau(T)}$ also depends on the  temperature.

In the next section, we present the thermodynamic origin of the Fourier and MCV equations, also discussing the embedding of temperature dependence into the thermodynamic parameters. Here, we exploit the second law of thermodynamics rigorously. 
In Section 3, the essential aspects of the numerical method  and its stability properties are discussed both for the Fourier and MCV equations in one spatial dimension. Then the effect of the parameters $a$ and $b$ is investigated through the numerical solutions.

\section{Nonlinear models of heat transport}
Let us consider a rigid heat conductor, and we recall the balance equation of internal energy $e$, which stands for the first law of thermodynamics:
\begin{equation}
\rho \partial_t e + \nabla\cdot\mathbf q = 0.
\end{equation}
Introducing the specific entropy  $s$ and the entropy flux density $\mathbf J_s$, the second law is expressed using the following inequality:
\begin{equation}\label{entropy}
\partial_t s+\nabla\cdot \mathbf J_s=\sigma_s \geq 0,
\end{equation}
where $\sigma_s$ is called entropy production. The solution of the inequality \eqref{entropy} is the constitutive equation, in our case that will be the Fourier or the MCV equations. In both cases, the entropy flux density $\mathbf J_s$ is
\begin{equation}
\mathbf J_s = \mathbf q/T.
\end{equation}
\subsection{Fourier heat equation}
In order to obtain the Fourier's law, one has to consider the local equilibrium hypothesis, i.e., $s=s(e)$ and the Gibbs relation,
\begin{equation}
T\textrm{d}s=\textrm{d}e,
\end{equation}
which leads to the inequality
\begin{equation}
\sigma_s=\mathbf q \cdot\nabla \Big(\frac{1}{T}\Big)\geq 0.
\end{equation}
Following Onsager, \cite{Ons}, we obtain
\begin{equation}\label{sigsolfou}
\mathbf q=l \nabla\Big(\frac{1}{T}\Big) = - \frac{l}{T^2}\nabla T = - \lambda(T) \nabla T,
\end{equation}
as a linear relationship between the thermodynamic flux and force. In \eqref{sigsolfou}, $l$ is the heat conduction coefficient and the nonlinearity is introduced through this parameter. It is apparent that the ratio $l/T^2$ is identified as being the thermal conductivity $\lambda$. In the linear case, $T^2$ is a reference temperature where the thermal conductivity is considered to be constant; thus the coefficient $l$ remains independent of the temperature. 
The expression of the  thermal conductivity  is compatible with the linear relation \eqref{ltdepT} if the relation 
\begin{equation} 
l=[\lambda_0 +a(T-T_0)]T^2
\end{equation}
holds, and $\lambda(T_0)=\lambda_0$. Finally, the following system forms the nonlinear Fourier heat equation:

\begin{subequations}
\begin{eqnarray}
&\rho \partial_t e + \nabla\cdot\mathbf q =0\label{nonlinearFourier},\\
&\mathbf q = - \lambda(T) \nabla T.
\end{eqnarray}
\end{subequations}

\subsection{Maxwell-Cattaneo-Vernotte equation}
As previously mentioned, the heat flux is a state variable now, i.e., $s=s(e, \mathbf q)$, and
\begin{equation}
s(e,\mathbf q)=s_{\textrm{eq}}(e)-\frac{m(e)}{2} \mathbf q\cdot \mathbf q,
\end{equation}
where $s_{\textrm{eq}}$ is the classical, local equilibrium part. The quadratic extension in the heat flux is the simplest one that preserves the convexity properties of entropy. For details, we refer to \cite{JouVasLeb88ext, CimJouRugVan, LebEtal08b, Gyar77a}. We also point out that the coefficient $m$ could be a function of $e$, for instance, which must be a positive definite function.
The entropy production is expressed as follows:
\begin{equation}
\sigma_s=\rho\frac{1}{T}\partial_t e-\rho m(e)\,\mathbf q\partial_t \mathbf q+\frac{1}{T}\nabla\cdot \mathbf q+\mathbf q\cdot \nabla\frac{1}{T}-\rho\frac{1}{2}\partial_e m(e) (\partial_t e)\mathbf q^2\geq 0.
\end{equation}
In the following we restrict ourselves to the case in which $\frac{\partial m(e)}{\partial e}=0$, that is, $m$ is a positive constant.
Thus, collecting the terms that are proportional to the heat flux $\mathbf q$, and substituting the internal energy balance \eqref{inten}, the entropy inequality becomes:
\begin{equation}\label{mcost}
\sigma_s=\left(-\rho m \partial_t \mathbf q+\nabla\left(\frac{1}{T}\right)\right)\cdot\mathbf q\geq 0.
\end{equation}
Theqrefore, as a solution of the above inequality, we obtain
\begin{equation}
-\rho m \partial_t \mathbf q+\nabla\left(\frac{1}{T}\right)= l\mathbf q.
\end{equation}
After rearrangement,
\begin{equation}
\frac{\rho m}{l} \partial_t \mathbf q+ \mathbf q=-\frac{1}{l\, T^2}\nabla T,
\end{equation}
we make the following identification:
\begin{equation}
\tau=\frac{\rho m}{l}, \quad \lambda=\frac{1}{l\, T^2},
\end{equation}
where it becomes apparent that the material parameters $\tau$ and $\lambda$ are not independent of each other. Moreover, the expression for $l$ is different than for Fourier's law, and it affects the definition of $\tau$ as well. 
In order to obtain the linear expressions for $\lambda$ and $\tau$ (see equations \eqref{ltdepT}-\eqref{ltdepT_a}), the following constraints arise:
\begin{subequations}
\begin{eqnarray}\label{lambda_tau}
&l(T)=\frac{1}{[\lambda_0 +a(T-T_0)]T^2}\\\label{lambda_tau_a}
&\rho m=\frac{[\tau_0+b(T-T_0)}{[\lambda_0 +a(T-T_0)]T^2}.
\end{eqnarray} 
\end{subequations}

Since $m$ is a constant, it is necessary to consider a temperature-dependent mass density ($\rho=\rho(T)$). Consequently, to preserve the thermodynamical compatibility, it is not possible to arbitrarily introduce any state variable dependence in the material parameters. Moreover, as $\rho=\rho(T)$, it refers to the presence of mechanical effects. In summary, if the temperature dependence of $\tau$ and $\lambda$ is present in an experiment, then the mechanical effects must be considered as well in the interpretation of the heat conduction process.

If $\frac{\partial m}{\partial e}\neq 0$, the entropy production includes  an additional  term as underlined. 
Theqrefore, considering the case in which the quantity $m$ is a function depending on the internal energy (and the temperature, respectively), the entropy production becomes:
\begin{equation}\label{mnocost}
\left(-\rho m(e) \partial_t \mathbf q+\nabla\left(\frac{1}{T}\right)-\frac{\rho}{2}\partial_e m(e)\,(\nabla\cdot\mathbf q )\,\,\mathbf q\right)\cdot\mathbf q\geq 0,
\end{equation}
and the following constitutive relation is obtained:
\begin{equation}\label{const_eq}
\mathbf q=-\frac{1}{2 l}\partial_e m(e)\,(\nabla\cdot\mathbf q)\,\mathbf q-\frac{\rho m(e)}{l}\partial_t \mathbf q\,-
\frac{1}{l T^2}\nabla T.
\end{equation}
By comparing  the  \eqref{const_eq} with the MCV heat conduction equation,  one has
\begin{equation}\label{gen_mcv}
\tau(T)\partial_t \mathbf q+\mathbf q\underline{\left(1+\frac{1}{2 l}\partial_e m(e)\,(\nabla\cdot \mathbf q)\right)}=-\lambda(T) \nabla T.
\end{equation}
We observe here a contribution of a volumetric effect that could have a mechanical origin. Assuming the simplest one-dimensional case and dividing the \eqref{gen_mcv} with the underlined expression, it `distorts' the original definition of $\tau$ and $\lambda$.
Furthermore, let us remark that  in Eqs.\eqref{lambda_tau}-\eqref{lambda_tau_a},  both $m$ and $\rho$ may depend on temperature. 

 We will not investigate (\ref{gen_mcv}); hence, in the following,   we refer to the case of a constant coefficient $m$. However, we fixed our attention to the fact that it is not trivial how to implement the nonlinear terms into the constitutive equations, and they influence other parameters. For example, the relations given by \eqref{lambda_tau}-\eqref{lambda_tau_a} may have  consequences on  mass density. In summary, we call the following system as the nonlinear MCV heat equation 
 \begin{subequations}
\begin{eqnarray}\label{nonlinearMCV}
&\rho(T) \partial_t e + \nabla\cdot\mathbf q = 0, \\ \label{nonlinearMCV_a}
&\tau(T)\partial_t \mathbf q+\mathbf q = - \lambda(T) \nabla T,
\end{eqnarray}
\end{subequations}
with emphasizing that possible mechanical effects would be more appropriate to include {such as thermal expansion. From this aspect, we would like to refer to the literature \cite{SellCimm19a, SellCimm19b, IgnOst10b}.} However, it is our intention only to investigate the above system of partial differential equations first, we do not expect rigorous physical interpretation from the solutions.

{In our last remark, we are mentioning a kind of `paradox' related to
the anomalous entropy production by the MCV equation, even in its linear case \cite{BarZan97, BarZan98, Zan99}, demonstrated on Taitel's problem. Here, the essential part is about the appropriate formulation of entropy production, and one cannot avoid the proper derivation. Here, in the nonlinear case, it becomes more important as it proposes further terms into the evolution equations. }

\section{Numerical aspects}
The system \eqref{nonlinearMCV}-\eqref{nonlinearMCV_a} may contain coefficients with several orders of magnitude difference. From numerical point of view, it is pronouncedly unfavorable. Thus it is convenient to introduce dimensionless parameters, following \cite{KovVan15, BCTFGGAGPV13},
\begin{equation}
\hat x = \frac{x}{L}, \quad \hat t = \frac{\alpha_0 t}{L^2}, \quad \alpha_0=\frac{\lambda(T_0)}{\rho_0 c_v}, \qquad\rho_0=\rho(T_0)
\end{equation}
 where $L$ is the length of the rigid and isotropic conductor, $T_0$ is the initial temperature, $\rho_0$ is the value of the mass density corresponding to the temperature $T_0$. The present combination of parameters particularly fits to the so-called flash or heat pulse experiment. This is a common, widely used methodology to measure the thermal material parameters either in low or room temperature situations \cite{ParEtal61, James80}. 
 
Exploiting   \eqref{lambda_tau_a}, it yields the following expressions:
\begin{equation}
\rho(T)=\frac{1}{m}\frac{\tau(T)}{\lambda(T)T^2}=\rho_0+\rho_1
\end{equation}
denoting 
\begin{equation}
\rho_0=\frac{1}{m}\frac{\tau_0}{\lambda_0T_0^2},\qquad
\rho_1=\frac{1}{m}\frac{b(T-T_0)}{[\lambda_0+a(T-T_0)]T^2},
\end{equation}
thus 
\begin{equation}
\rho_1=\rho_0 \frac{b(T-T_0)}{\tau_0}.
\end{equation}
Moreover
\begin{equation}
\lambda(T_0)=\frac{1}{l(T_0) T_0^2},
 \qquad \hat{T} = \frac{T-T_0}{T_{\hbox{end}}-T_0}
\end{equation}
with $l(T_0)$ being the phenomenologial coefficient at $T_0$ and $ T_{\hbox{end}}$ is the equilibrium temperature corresponding to adiabatic boundaries, i.e.,
\begin{subequations}
\begin{eqnarray}
&T_{\hbox{end}} = T_0 + \frac{1}{\rho c_v L} \int_{t_0}^{t_p} q_0 (t) \, \hbox{d}t, \\
& \bar q_0 = \frac{1}{t_p} \int_{t_0}^{t_p} q_0 (t) \, \hbox{d}t, \quad \hat q = \frac{q}{\bar q_0},
\end{eqnarray}
\end{subequations}
where $t_p$ is the length of the pulse that acts on the boundary as a heat pulse. Moreover, $t_0$ is the initial time instant which is considered to be $0$ and $\bar q_0$ is the integral average of the heat pulse $q_0(t)$. In order to simplify the notations, the `hat' is omitted in the following and only dimensionless parameters are used. 

Now, in 1D, the dimensionless system of equations reads 
\begin{subequations}
\begin{eqnarray}
&\left(\tau_{p_1} +\tau_{p_3} T\right)\partial_t T + \partial_x q =0,  \\
&(\tau_{q_1} + \tau_{q_2} T) \partial_t q + q = - (\tau_{p_1} + \tau_{p_2} T) \partial_x T,
\end{eqnarray}
\end{subequations}
with 
\begin{subequations}
\begin{eqnarray}
&\tau_{p_1}=\frac{\alpha_0 t_p}{L^2},\qquad
 \tau_{p_2}=\frac{a (T_{\textrm{end}}-T_0) t_p}{\rho c_v L^2},\qquad
\tau_{p_3}=\tau_{p_1}\frac{\tau_{q_2}}{\tau_{q_1}},\\
& \tau_{q_1}=\frac{\alpha_0 \tau_0}{L^2}, \qquad \tau_{q_2}=\frac{\alpha_0 b (T_{\textrm{end}}-T_0)}{L^2}.
\end{eqnarray}
\end{subequations}
\subsection{Difference equations}
\begin{figure}[H]
\centering
\includegraphics[width=8cm,height=1.5cm]{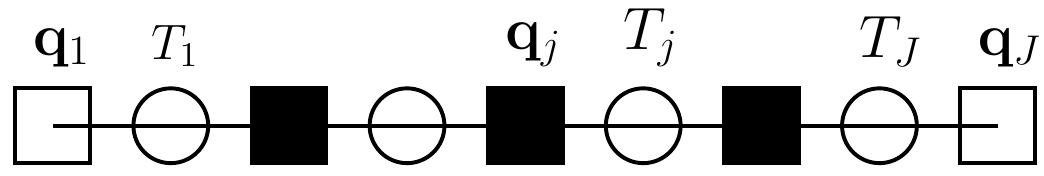}
\caption{Concept of the spatial discretization \cite{RietEtal18}.}
\label{FIG1}
\end{figure}

The present numerical method is developed first for linear heat equations and validated using an analytical solution for the Guyer-Krumhansl equation \cite{Kov18gk}. The fundamental principle remains the same: we use a staggered spatial field to distinguish `surface' and `volume average' quantities. In heat pulse experiments, when the heat flux is defined on both boundaries, the temperature is shifted by half space step $(\Delta x)/2$, as it is shown in Fig.~\ref{FIG1}. As a consequence, there is no need to define boundaries for the temperature field. In time,
Theqrefore the difference equations are only an explicit forward differencing scheme is used.

\begin{subequations}
\begin{eqnarray}
&\frac{\tau_{p_1}}{\Delta t} \left (1+\frac{\tau_{q_2}}{\tau_{q_1}} T^n_j\right) \left(T^{n+1}_{j}-T^n_j \right )  = -\frac{1}{\Delta x} \left ( q^n_{j+1}- q^n_j \right ),  \label{DMCV}
\\\label{DMCV_a}
&\frac{\tau_{q_1} + \tau_{q_2} T^n_j}{\Delta t}\left ( q^{n+1}_{j}-q^n_j \right )  = -q^n_j - \frac{\tau_{p_1} + \tau_{p_2} T^n_j}{\Delta x} \left ( T^n_{j}- T^n_{j-1} \right ),
\end{eqnarray}
\end{subequations}
where $n$ denotes the time steps and $j$ stands for the spatial steps.

\subsection{Stability analysis}
\subsubsection{I. Fourier heat equation}
First of all, let us consider only the Fourier heat equation, i.e., $\tau_{q_1}=\tau_{q_2}=0$ in \eqref{DMCV_a}. It is important to emphasize that the conventional analysis using the von Neumann method is not applicable directly as the stability conditions themselves depend on the temperature, at least in the present case\footnote{That kind of dependency is up to the choi1ce of nonlinearities.}. 
The paper of Weickert et al.~\cite{WeickEtal98} proposes a stability analysis method for nonlinear diffusion equations, and demonstrates on particular examples from image processing \cite{Weick96a, Weick97rev}. In order to utilize this approach, one must reformulate the difference equations in a way to obtain a mapping between two time instants of temperature: $T^{n} \rightarrow T^{n+1}$. That mapping is represented by a tridiagonal matrix $\mathbf Q$ with elements $q_{ij}$, and it must satisfies the following criteria:
\begin{itemize}
\item continuity in the $T$-dependence,
\item symmetry, $q_{ij}=q_{ji}$,
\item unit row sum, $\sum_i q_{ij} = 1$,
\item non-negativity, $q_{ij}>0$,
\item positive diagonal elements, $q_{ii}>0$,
\item irreducibility, i.e., for any $T\geq0$, $\lambda(T)>0$.
\end{itemize}
The corresponding tridiagonal matrix has the elements $[ \beta; 1-2\beta; \beta]$ in a row with $\beta=\Delta t (\Delta x)^2 (\tau_{p_1} + \tau_{p_2} T^n_j)/\tau_{p_1}>0$, any other element is zero. Hence the non-negativity and positive diagonality requirements are equivalent and reads as 
\begin{equation}
1> 2 \max_j \beta,
\end{equation}
that is, one must estimate the maximum value of the temperature field. In the simulation of a heat pulse experiments using the Fourier's law, it is simple as the equilibrium dimensionless temperature is $1$ and it cannot be higher due to the particular dimensionless formulation of temperature. 
It restricts the maximum time step:
\begin{equation}
\Delta t < \frac{(\Delta x)^2 \tau_{p_1}}{2 (\tau_{p_1} + \tau_{p_2})}.
\end{equation}

As an alternative way, one could assume apriori the maximum temperature and applying a linear stability analysis by following von Neumann's method \cite{NumRec07b} and Jury conditions \cite{Jury74}. Briefly speaking, it starts with assuming a solution in a plane wave form:
\begin{equation}\label{wave}
\phi^n_j = \xi^n e^{ikj\Delta x},
\end{equation}
where $i$, $k$ and $\xi$ are the imaginary unit, wave number and growth factor, respectively; the stability condition is $|\xi|\leq 1$, i.e., the amplitude of the wave remains bounded. After substituting it into the difference equations, it yields a characteristic polynomial for $\xi$: $F(\xi)=a_0 + a_1 \xi + a_2 \xi^2$, with 
\begin{subequations}
\begin{eqnarray}
&a_0= 0,  \\
&a_1 =\frac{4 \Delta t (\tau_{p_1}+\tau_{p_2})}{(\Delta x)^2 \tau_{p_1}} - 1, \\
&a_2=1.
\end{eqnarray}
\end{subequations}
Then applying the Jury conditions in order to restrict the roots of $F(\xi)$ to keep them inside the unit circle on the complex plane, we obtain
\begin{enumerate}
\item $F(\xi=1) \geq 0$, i.e., $a_0 + a_1 + a_2 \geq 0$, which is automatically fulfilled,
\item $F(\xi=-1) \geq 0$, i.e., $a_0 - a_1 + a_2 \geq 0$ that yields the same stability criterion as the other method, and 
\item $|a_0| \leq 1$ which is trivially satisfied.
\end{enumerate}
Thus it is possible to apply the linear stability analysis if one can apriori estimate the upper bound of temperature. As it is not proved for other nonlinearities such as thermal radiation, it is safe to state that it is true only if the nonlinearity occurs due to the temperature dependence in the thermal conductivity. 

\subsubsection{ II. MCV heat equation}
While Weickert et al.~\cite{WeickEtal98} assumed only one field variable, which case would refer to the elimination of heat flux. We now recall our strategy: it is not our intention to eliminate any variables as the outcome would be unreasonably difficult to handle. Instead, we use both variables in the following.

Before we proceed with the previously presented stability analysis, let us estimate directly the stability criterion for the nonlinear MCV equation. In case of the Fourier heat equation, we have seen that the $\tau_{p_1}/(\tau_{p_1}+\tau_{p_2})$ correction appears in the stability condition that lowers the maximum allowable time step in the algorithm, comparing to the linear case. Analogously, we assume that a similar correction, $\tau_{q_1}/(\tau_{q_1} + \tau_{q_2})$ appears as well. Using the results of the linear case \cite{RietEtal18}, which is
\begin{equation}\label{delta_t}
\Delta t < \frac{(\Delta x)^2}{4},
\end{equation}
we expect that
\begin{equation}
\Delta t < \frac{(\Delta x)^2}{4} \frac{\tau_{p_1}}{\tau_{p_1}+\tau_{p_2}\max_{(j,n)} T^n_j} \label{STABMCVNL}
\end{equation}
will appear without the relaxation time as it does not appear even in the linear case. However, with the estimation for temperature $\max_{(j,n)} T^n_j=1$, it could be too optimistic as the MCV model is a damped wave equation. It is safe to say that the $\max_{(j,n)} T^n_j\approx 3$ for real parameters that could occur in experiments. This approximation means that the maximum temperature in the simulation is 3 times higher than the equilibrium one. We must note that this maximum strongly depends on $\tau_{q_1}$ and higher temperature values could occur. 
Let us assume the $\max_{(j,n)} T^n_j=3$ apriori.

Following the same procedure of von Neumann,
assuming a solution of the difference equation in the form \eqref{wave},
 after substituting it into the difference equations,
we obtain the system of linear algebraic equations:
\begin{equation}
\mathbf M\cdot
\left( \begin{array}{l}
T_0\\ q_0
\end{array}\right )
=0
\end{equation}
where the coeffcient matrix is given:
\begin{equation}
\mathbf M=
\left(
\begin{array}{ll}
\frac{\tau_{q_1}}{\Delta t}\left(1+ \frac{\tau_{q_2} C}
{\tau_{q_1}}\right)(\xi-1)&
\frac{1}{\Delta x}
\left(e^{ik\Delta x}-1\right)\\
\left(1-e^{-ik\Delta x}\right)\frac{\tau_{p_1}+\tau_{p_2} C}{\Delta x}&
1+\frac{\tau_{q_1}+\tau_{q_2}C}{\Delta t}\left(\xi-1\right)
\end{array}
\right)
\end{equation}
with $C$ being the maximum value for $T^n_j$.

 The characteristic polynomial for $\xi$
 ($\det\mathbf M=0$) can be expressed as
\begin{equation}
F(\xi)=a_0 + a_1 \xi + a_2 \xi^2,
\end{equation} 
with 
\begin{subequations}
\begin{eqnarray}
&a_0=-\frac{\tau_{p_1}\tau_q}{\tau_{q_1}\Delta t}
+\frac{\tau_q^2 \tau_{p_1}}{(\Delta t)^2\tau_{q_1}}-\frac{2}{(\Delta x)^2}\left(\tau_{p_1}+\tau_{p_2} C\right)\left(\cos(k\Delta x)-1\right),\\
&a_1=\frac{\tau_{p_1}\tau_{q}}{\tau_{q_1}\Delta t}-\frac{2 \tau_{q}^2\tau_{p_1}}{(\Delta t)^2\tau_q},\\
&a_2=\frac{\tau_q \tau_{p_1}}{(\Delta t)^2\tau_{q_1}},
\end{eqnarray}
\end{subequations}
wherein $\tau_q=(\tau_{q_1}+\tau_{q_2}C)$.
Then applying again the same  Jury criteria, it yelds the conditions below,
\begin{enumerate}
\item $F(\xi=1) \geq 0$, i.e., $a_0 + a_1 + a_2 \geq 0$, i.e.,
\begin{equation}
\frac{4(\tau_{p_1}+\tau_{p_2}C)}{(\Delta x)^2}\geq 0
\end{equation}
which is trivially satisfied.
\item $F(\xi=-1) \geq 0$, i.e., $a_0 - a_1 + a_2 \geq 0$ if the condition
\begin{equation}
\frac{\tau_q}{\tau_{q_1}\Delta t}\left[\frac{\tau_q}{\Delta t}-\frac{1}{2}
\right]+\frac{(\tau_{p_1}+\tau_{p_2}C)}{(\Delta x)^2\tau_{p_1}}\geq 0
\end{equation}
holds.
\item $|a_0| \leq a_2$ which is satisfied if the following inequality:
\begin{equation}\label{ineq}
\Delta t \leq \frac{\Delta x^2}{4}\frac{\tau_{p_1}}{\tau_{p_1}+\tau_{p_2}C}\frac{\tau_{q_1}+\tau_{q_2}C}{\tau_{q_1}}.
\end{equation}

\end{enumerate}
Let us remark that in linear case, i.e. for $\tau_{q_2}=0, \tau_{p_2}=0$, the inequality \eqref{ineq} reduces to the condition \eqref{delta_t}. Furthermore, the correction related to the relaxation time also appears in the stability condition. We note here that this is the consequence of the temperature dependence of mass density.


\section{Results and discussion}
In this final section, we demonstrate the solutions of nonlinear Fourier and MCV heat equations and investigate the effects of nonlinear terms. Here, only the rear side ($\hat x=1$) temperature histories are presented because, in the heat pulse experiments, this one is measured and used for evaluation. 

\subsection{\bf{Initial and boundary conditions}}
In such a measurement setup, it is required to have homogeneous temperature distribution at the beginning, and the sample must be in thermal equilibrium with its environment. Thus both fields are zero at the initial time instant. 
Regarding the boundary conditions, the heat pulse excites the front side of the sample. For numerical reasons, it is favorable to use a smooth function. Its dimensionless form reads as
\begin{center}
	$q( x=0, t)= \left\{ \begin{array}{cc}
	 \left(1-\cos\left(2 \pi \cdot \frac{ t}{t_p}\right)\right) &
	\textrm{if } 0<  t \leq  t_p,\\
	0 & \textrm{if }  t>  t_p,
            \end{array} \right.  $
\end{center}
and the rear side is considered to be adiabatic, i.e, $q(x=1,t)=0$.

\subsection{Solutions}
\begin{figure}
\centering
\includegraphics[width=12cm,height=6cm]{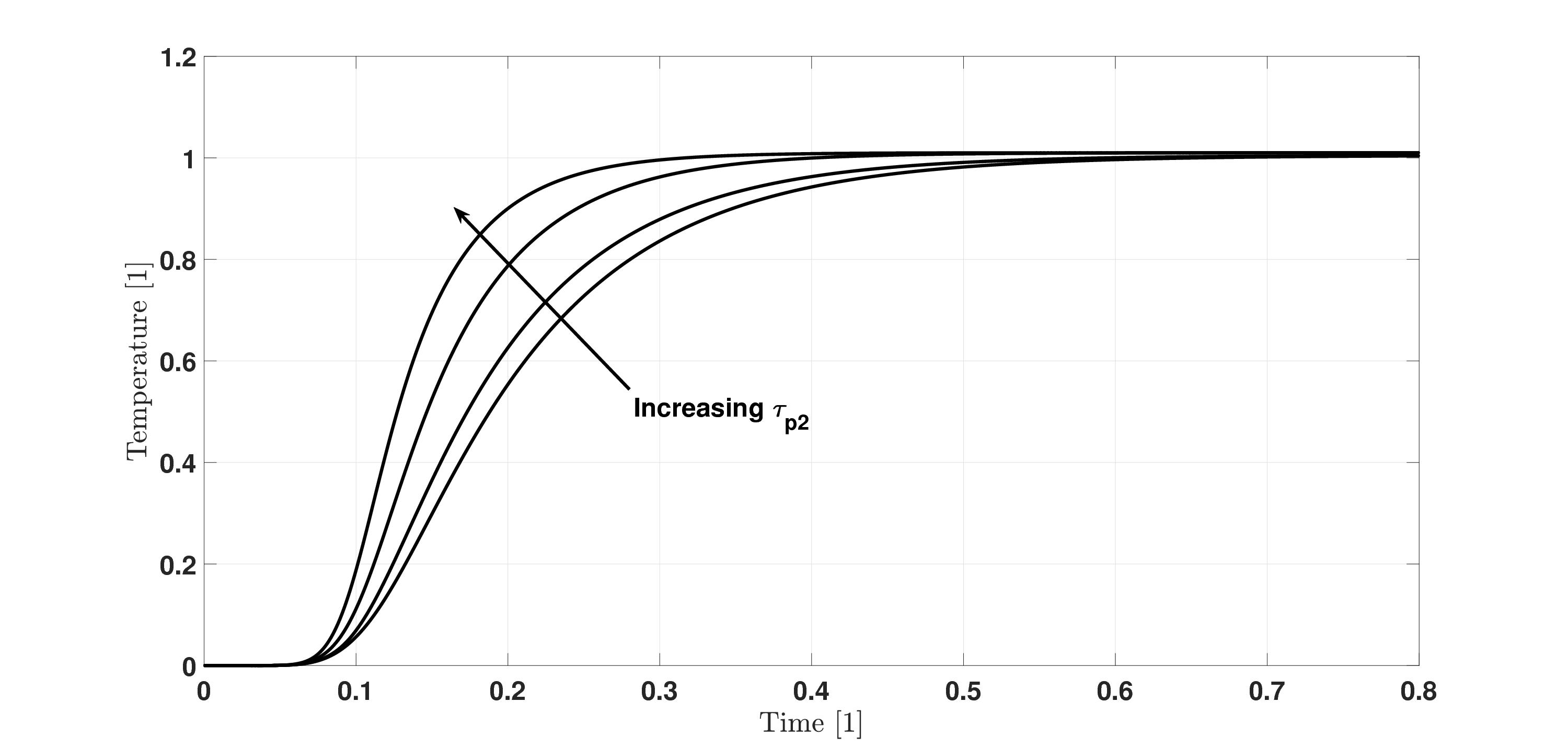}
\caption{The rear side temperature history, using the Fourier heat equation with increasing $\tau_{p_2}$ parameters: $0$, $0.01$, $0.05$, $0.1$, respectively; $\tau_{p_1}=0.1$.}
\label{fig1}
\end{figure}

\begin{figure}
\centering
\includegraphics[width=12cm,height=6cm]{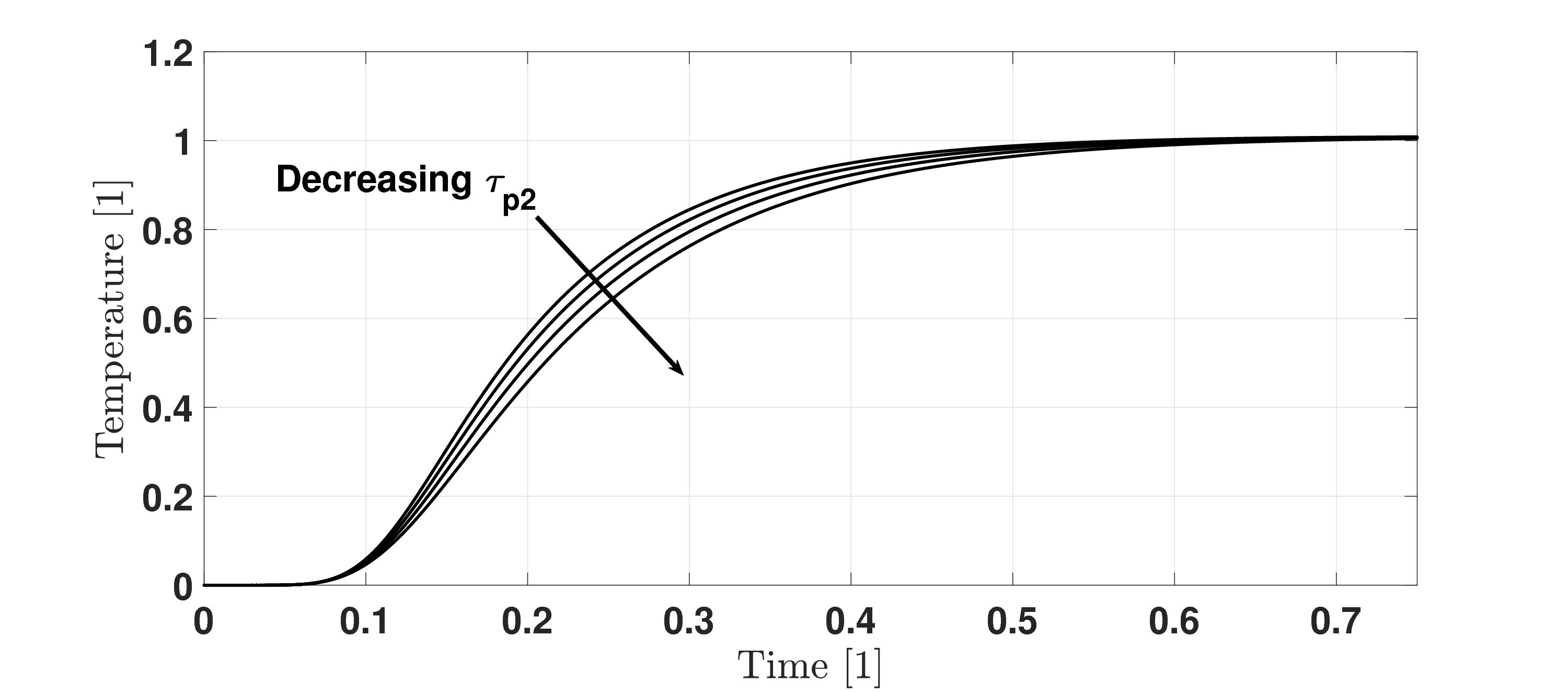}
\caption{The rear side temperature history, using the Fourier heat equation with increasing $\tau_{p_2}$ parameters: $0$, $-0.005$, $-0.01$, $-0.015$, respectively; $\tau_{p_1}=0.1$.}
\label{fig2f}
\end{figure}

Regarding the nonlinear Fourier equation (see Figs.~\ref{fig1} and \ref{fig2f}), the $\tau_{p_2}$ parameter influences the slope at the point when the temperature starts to increase. It is worth to observe that the point corresponding to $\hat T=0.5$ is significantly shifted to the left for increasing $\tau_{p_2}$. It is important because the conventional evaluation formula for the Fourier heat equation, which offers the thermal diffusivity as an outcome of the measurement, uses the time instant related to $\hat T=0.5$. {Here we emphasize that negative coefficient for the temperature dependence is also physically admissible and possible in several practical cases. However, one has to pay attention that the thermal conductivity must remain positive, restricted by the second law. Thus we are also testing the solutions for negative $\tau_{p_2}$, see Fig.~\ref{fig2f}. It affects the slope oppositely.}

\begin{figure}
\centering
\includegraphics[width=12cm,height=6cm]{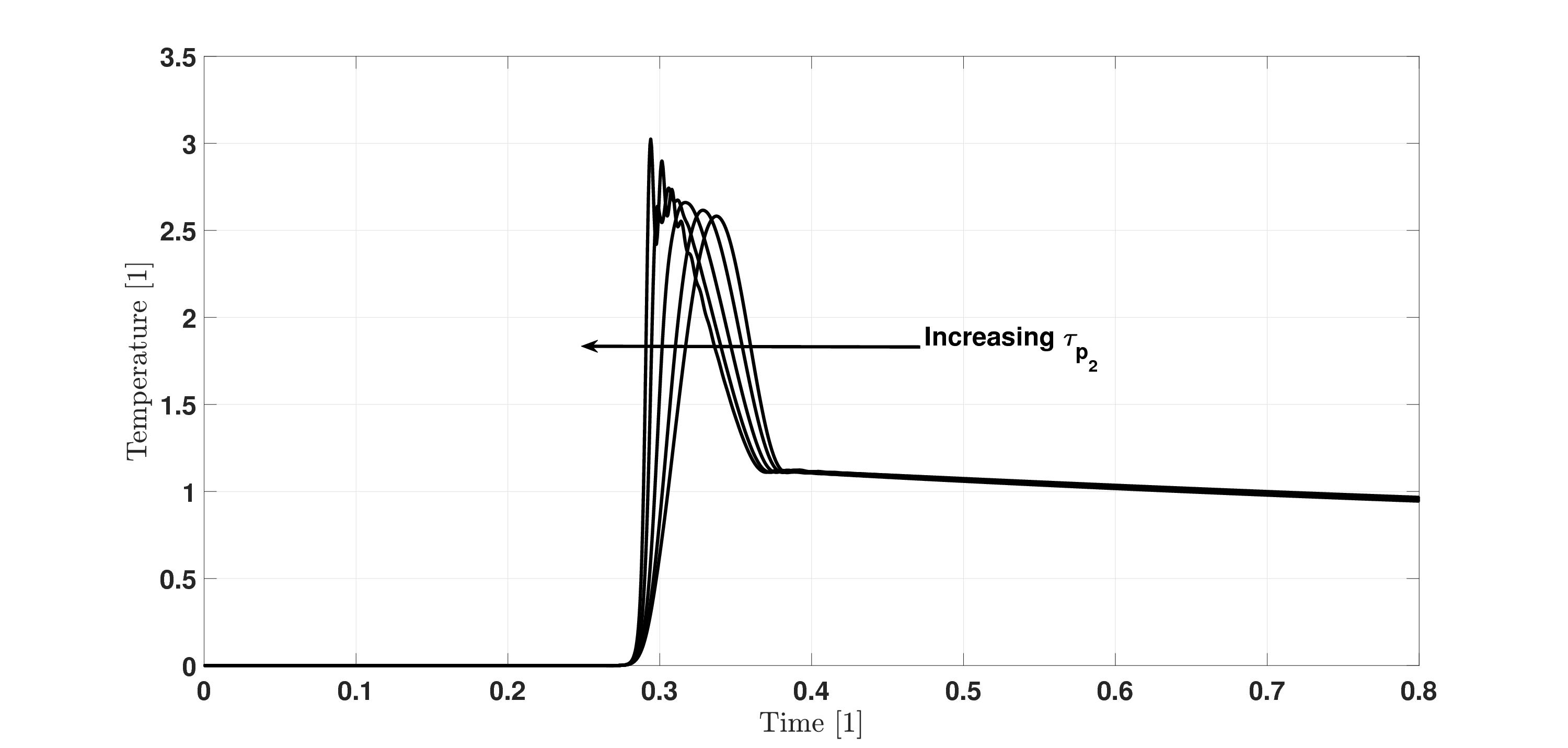}
\caption{The rear side temperature history, using the MCV heat equation with increasing $\tau_{p_2}$ parameters: $0$, $0.001$, $0.002$, $0.005$, $0.01$, respectively; $\tau_{p_1}=0.1$, $\tau_{q_1}=0.08$, $\tau_{q_2}=0$.}
\label{fig2}
\end{figure}

\begin{figure}
\centering
\includegraphics[width=12cm,height=6cm]{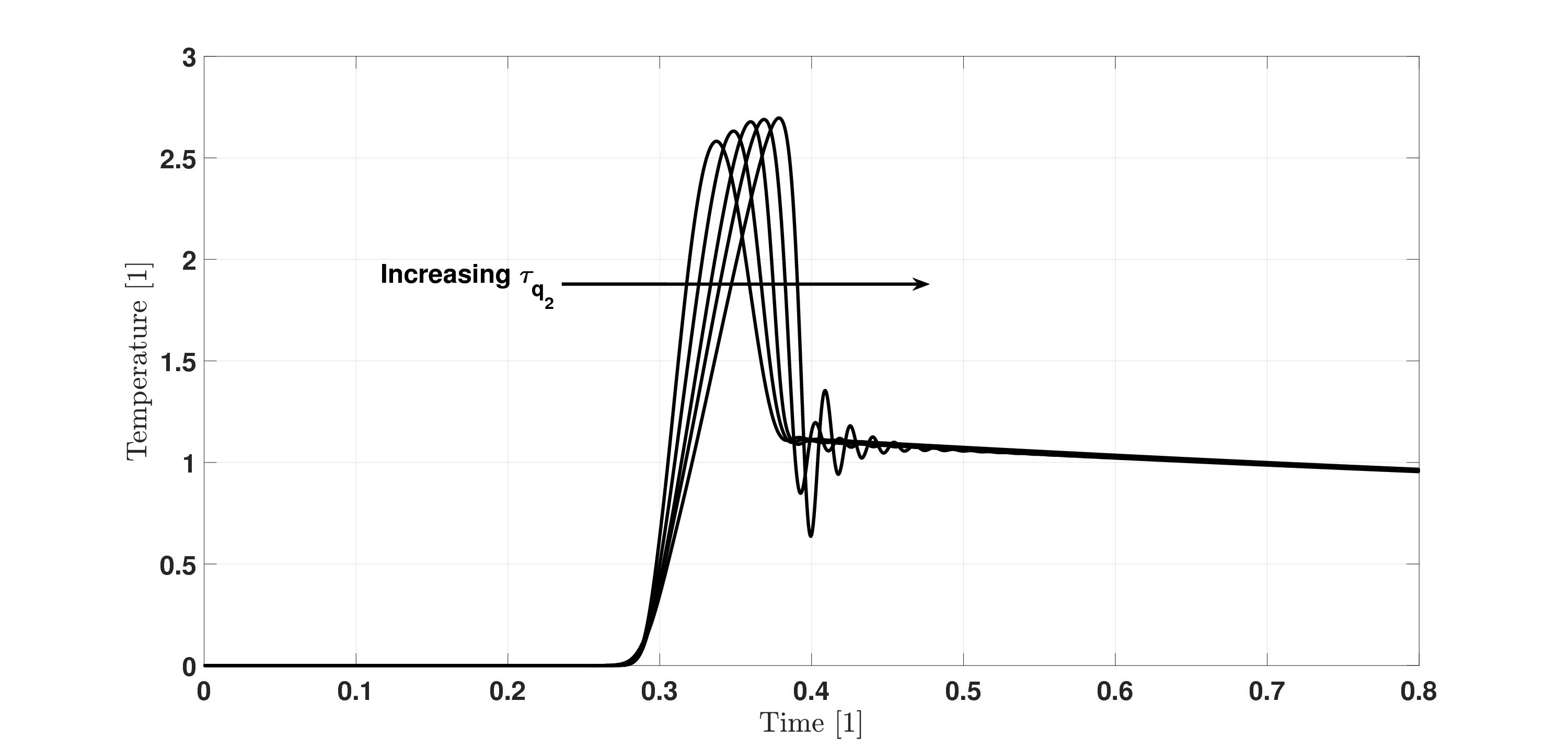}
\caption{The rear side temperature history, using the MCV heat equation with increasing $\tau_{q_2}$ parameters: $0$, $0.001$, $0.002$, $0.003$, $0.004$, respectively; $\tau_{p_1}=0.1$, $\tau_{p_2}=0$, $\tau_{q_1}=0.08$.}
\label{fig3}
\end{figure}

Investigating the effects of the same parameter in the solutions of the MCV equation, we experience similar effects (see Fig.~\ref{fig2}). Using $\tau_{p_1}=0.1$, $\tau_{q_1}=0.08$ parameters, the wave signal dominates the solution, and that characteristic is pushed to the left again, affecting only the slope of the wave edge. However, as one can observe, the MCV solution becomes dispersive for larger $\tau_{p_2}$. That property of the presented scheme holds for the other cases in which the $\tau_{q_2}$ is increased (see Fig.~\ref{fig3}). Seemingly, these parameters act against each other, $\tau_{q_2}$ shifts the wave signal to the right. In both situations, the solution remains stable, and the dispersive error can be decreased by increasing the resolution of the discretization. 

\begin{figure}
\centering
\includegraphics[width=12cm,height=6cm]{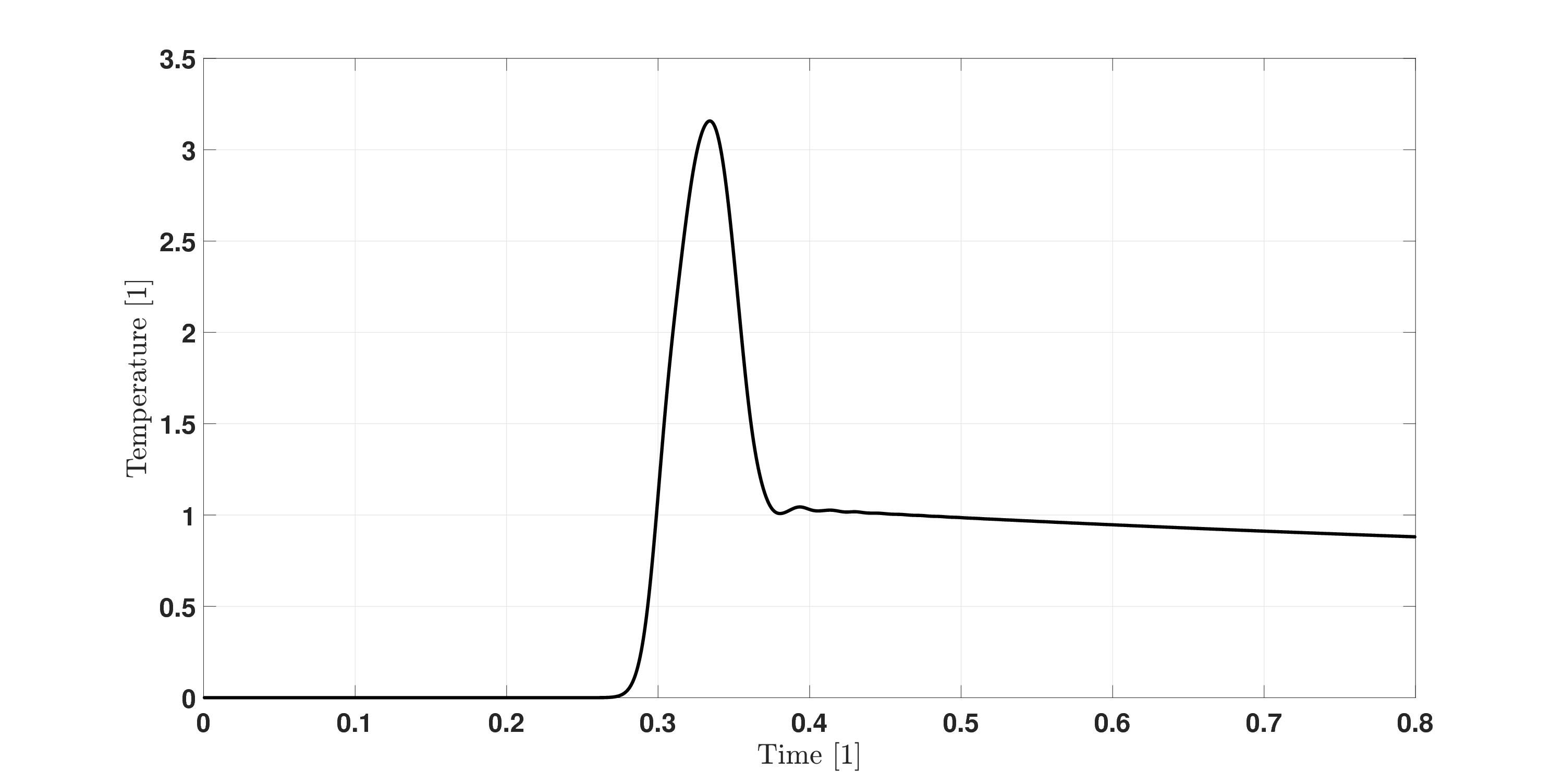}
\caption{The rear side temperature history, using the MCV heat equation, using $\tau_{p_1}=0.1$, $\tau_{p_2}=0.03$, $\tau_{q_1}=0.08$, $\tau_{q_2}=0.01$.}
\label{fig4}
\end{figure}

Despite the simplicity of the scheme, it is able to solve both the nonlinear Fourier and MCV heat equations. The dispersive errors are present only when one of the parameters corresponding to the temperature dependence begins to dominate. Let us remark that  Fig.~\ref{fig4} demonstrates a different case in which both parameters ($\tau_{p_2}$ and $\tau_{q_2}$) are much larger than previously and the numerical solutions becomes free from artificial oscillations, using the same discretization, too. It is because these parameters affect the solution oppositely. 

\section{Summary}

In the present paper, we first discussed the thermodynamic origin of the Fourier and MCV equations. This derivation is based only on the first and second laws of thermodynamics, showing the connection between the material parameters. It becomes exceptionally crucial for non-Fourier heat conduction and could be more challenging to deal with for further extensions beyond the MCV equation. Even for the MCV heat equation, it turned out that the simplest -- linear -- temperature dependence both in relaxation time and thermal conductivity induces a temperature-dependent mass density, not independently of the other material parameters. Thus, strictly speaking, the mechanical effects induced by $\rho(T)$ could not be negligible. {It is our intention to include possible mechanical effects such as thermal expansion in the future work to obtain a proper system of equations, describing a more realistic situation.}

However, our goal was only to demonstrate how a numerical scheme, using staggered fields, operates for nonlinear problems. We also concluded that the nonlinear stability analysis might not be necessary for these special, but practically important nonlinearities. Instead, we applied the method of von Neumann for stability analysis, apriori assuming the maximum temperature. 

We performed a parameter sweep to show the effect of temperature-dependent coefficients. First, we observed that $\tau_{p_2}$ and $\tau_{q_2}$ affects the solution oppositely. Secondly, their effect is detectable by measuring the slope at the foot of the wave signal. It could be important for experimental analysis. 

\section{Acknowledgement}
The research reported in this paper was supported by the Higher Education Excellence Program of the Ministry of Human Capacities in the frame of Nanotechnology research area of Budapest University of Technology and Economics (BME FIKP-NANO).
The work was supported by the grants of National Research, Development and Innovation Office – NKFIH 124366, NKFIH 123815, NKFIH 124508 and NKFIH 130378, and by FIEK-16-1-2016-0007. The research reported in this paper has been supported by the National Research, Development and Innovation Fund (TUDFO/51757/2019-ITM), Thematic Excellence Program.

The authors acknowledge the financial support of the Italian Gruppo Nazionale per la Fisica Matematica (GNFM-INdAM).


\begin{thebibliography}{10}


\bibitem{Tisza38}
L.~Tisza.
\newblock {Transport phenomena in {H}elium {II}}.
\newblock {\em Nature}, 141:913, 1938.

\bibitem{Pesh44}
V.~Peshkov.
\newblock Second sound in {H}elium {II}.
\newblock {\em J. Phys. (Moscow)}, 381(8), 1944.

\bibitem{JacWalMcN70}
H.~E. Jackson, C.~T. Walker, and T.~F. McNelly.
\newblock Second sound in {N}a{F}.
\newblock {\em Physical Review Letters}, 25(1):26--28, 1970.

\bibitem{Gyarmati70b}
I.~Gyarmati.
\newblock {\em Non-equilibrium thermodynamics}.
\newblock Springer, 1970.

\bibitem{FulEta14m1}
T.~Fülöp, Cs. Asszonyi, and P.~V\'an.
\newblock Distinguished rheological models in the framework of a
  thermodynamical internal variable theory.
\newblock {\em Continuum Mechanics and Thermodynamics}, 27(6):971--986, 2015.

\bibitem{SzucsFul18a}
M~Szücs and T.~Fülöp.
\newblock Kluitenberg-{V}erh\'as rheology of solids in the {GENERIC} framework.
\newblock  {\em Journal of Non-Equilibrium Thermodynamics}, 44(3):247-259, 2019.

\bibitem{SzucsFul18b}
M~Szücs and T.~Fülöp.
\newblock Analytical solution method for rheological problems of solids.
\newblock 2018.

\bibitem{CarMorr72a}
M.~Carrasi and A.~Morro.
\newblock A modified {N}avier-{S}tokes equation, and its consequences on sound
  dispersion.
\newblock {\em Il Nuovo Cimento B}, (9):321--343, 1972.

\bibitem{LebCloo89}
G.~Lebon and A.~Cloot.
\newblock Propagation of ultrasonic sound waves in dissipative dilute gases and
  extended irreversible thermodynamics.
\newblock {\em Wave Motion}, (11):23--32, 1989.

\bibitem{SellEtal16b}
A.~Sellitto, V.~A. Cimmelli, and D.~Jou.
\newblock {\em Mesoscopic theories of heat transport in nanosystems}.
\newblock Springer, Berlin, 2016.

\bibitem{Verhas97}
J.~Verhás.
\newblock {\em Thermodynamics and {R}heology}.
\newblock Akad\'emiai Kiad\'o-Kluwer Academic Publisher, 1997.

\bibitem{Str11b}
B.~Straughan.
\newblock {\em Heat waves}.
\newblock Springer, Berlin, 2011.


\bibitem{Max1867}
J.~C. Maxwell.
\newblock On the dynamical theory of gases.
\newblock {\em Philosophical Transactions of the Royal Society of London},
  157:49--88, 1867.

\bibitem{Cattaneo58}
C.~Cattaneo.
\newblock Sur une forme de lequation de la chaleur eliminant le paradoxe dune
  propagation instantanee.
\newblock {\em Comptes Rendus Hebdomadaires Des Seances De L'Academie Des
  Sciences}, 247(4):431--433, 1958.


\bibitem{Cat1}
C. Cattaneo.
 \newblock Sulla conduzione del calore.
\newblock{Atti Sem. Mat. Fis. Univ. Modena}, 247, 83-101,1948.


\bibitem{Vernotte58}
P.~Vernotte.
\newblock Les paradoxes de la th{\'e}orie continue de l{\'e}quation de la
  chaleur.
\newblock {\em Comptes Rendus Hebdomadaires Des Seances De L'Academie Des
  Sciences}, 246(22):3154--3155, 1958.


\bibitem{Gyar77a}
I.~Gyarmati.
\newblock On the wave approach of thermodynamics and some problems of
  non-linear theories.
\newblock {\em Journal of Non-Equilibrium Thermodynamics}, 2:233--260, 1977.


\bibitem{JouVasLeb88ext}
D.~Jou, J.~Casas-V{\'a}zquez, and G.~Lebon.
\newblock Extended {I}rreversible {T}hermodynamics.
\newblock {\em Reports on Progress in Physics}, 51(8):1105, 1988.

\bibitem{LebEta11a}
G.~Lebon, M.~Hatim, M.~Grmela, and Ch. Dubois.
\newblock An extended thermodynamic model of transient heat conduction at
  sub-continuum scales.
\newblock 467(2135):3241--3256, 2011.

\bibitem{LebEtal08b}
G.~Lebon, D.~Jou, and J.~Casas-V{\'a}zquez.
\newblock {\em Understanding {N}on-equilibrium {T}hermodynamics}.
\newblock Springer, 2008.

\bibitem{JouEtal99}
D.~Jou, J.~Casas-Vazquez, and G.~Lebon.
\newblock Extended irreversible thermodynamics revisited (1988-98).
\newblock {\em Reports on Progress in Physics}, 62(7):1035, 1999.

\bibitem{Lebon14}
G.~Lebon.
\newblock Heat conduction at micro and nanoscales: a review through the prism
  of extended irreversible thermodynamics.
\newblock {\em Journal of Non-Equilibrium Thermodynamics}, 39(1):35--59, 2014.


\bibitem{CimKovVanRog}
V.A. Cimmelli, R. Kovács, P. V\'an, P. Rogolino.
\newblock Generalized heat-transport equations: parabolic and hyperbolic models.
\newblock {\em Continuum Mechanics and Thermodynamics},    30(6),  1245-1258, 2018




\bibitem{RogCim}
P. Rogolino, V.A. Cimmelli.
\newblock{ Differential  consequences of balance laws in Extended Irreversible Thermodynamics of rigid heat conductors}, \newblock{\em Proceedings of the Royal of London A}, 475, 2227, 2019.


\bibitem{FairLaWill47}
C.~T. Lane, H.~A. Fairbank, and W.~M. Fairbank.
\newblock Second sound in liquid {H}elium {II}.
\newblock {\em Physical Review}, 71:600--605, 1947.



\bibitem{McN74t}
T.~F. McNelly.
\newblock Second {S}ound and {A}nharmonic {P}rocesses in {I}sotopically {P}ure
  {A}lkali-{H}alides.
\newblock 1974.
\newblock Ph.D. Thesis, Cornell University.


\bibitem{NarDyn72a}
V.~Narayanamurti and R.~C. Dynes.
\newblock Observation of second sound in bismuth.
\newblock {\em Physical Review Letters}, 28(22):1461--1465, 1972.

\bibitem{Naretal75}
V.~Narayanamurti, R.~C. Dynes, and K.~Andres.
\newblock Propagation of sound and second sound using heat pulses.
\newblock {\em Physical Review B}, 11(7):2500--2524, 1975.

\bibitem{Cimmelli09nl}
V.A. Cimmelli, A.~Sellitto, and D.~Jou.
\newblock Nonlocal effects and second sound in a non-equilibrium steady state.
\newblock {\em Physical Review B}, 79(1):014303, 2009.





\bibitem{GuyKru2}
  R. A. Guyer and J. A. Krumhansl.
\newblock{Thermal conductivity, second sound and phonon hydrodynamic phenomena in nonmetallic crystals},
   \newblock{\em Phys. Rev.},148, 778-788, 1966




\bibitem{GuyKru66a1}
R.~A. Guyer and J.~A. Krumhansl.
\newblock Solution of the linearized phonon {B}oltzmann equation.
\newblock {\em Physical Review}, 148(2):766--778, 1966.





\bibitem{GuyKru66a2}
R.~A. Guyer and J.~A. Krumhansl.
\newblock Thermal conductivity, second sound and phonon hydrodynamic phenomena
  in nonmetallic crystals.
\newblock {\em Physical Review}, 148(2):778--788, 1966.







\bibitem{KovVan15}
R.~Kovács and P.~Ván.
\newblock Generalized heat conduction in heat pulse experiments.
\newblock {\em International Journal of Heat and Mass Transfer}, 83:613 -- 620,
  2015.







\bibitem{BerVan17b}
A.~Berezovski and V\'an P.
\newblock {\em Internal Variables in Thermoelasticity}.
\newblock Springer, 2017.







\bibitem{Nyiri89}
B.~Nyíri.
\newblock On the extension of the {G}overning {P}rinciple of {D}issipative
  {P}rocesses to nonlinear constitutive equations.
\newblock {\em Acta Physica Hungarica}, 66(1):19--28, 1989.







\bibitem{Nyiri91}
B.~Nyí­ri.
\newblock On the entropy current.
\newblock {\em Journal of Non-Equilibrium Thermodynamics}, 16(2):179--186,
  1991.

\bibitem{Botetal16}
S.~Both, B.~Cz{\'e}l, T.~F{\"u}l{\"o}p, Gy. Gr{\'o}f, {\'A}.~Gyenis,
  R.~Kov{\'a}cs, P.~V{\'a}n, and J.~Verh{\'a}s.
\newblock Deviation from the {F}ourier law in room-temperature heat pulse
  experiments.
\newblock {\em Journal of Non-Equilibrium Thermodynamics}, 41(1):41--48, 2016.

\bibitem{Vanetal17}
P.~V{\'a}n, A.~Berezovski, T.~F{\"u}l{\"o}p, Gy. Gr{\'o}f, R.~Kov{\'a}cs,
  {\'A}.~Lovas, and J.~Verh{\'a}s.
\newblock Guyer-{K}rumhansl-type heat conduction at room temperature.
\newblock {\em EPL}, 118(5):50005, 2017.
\newblock arXiv:1704.00341v1.



\bibitem{RietEtal18}
Á. Rieth, R.~Kovács, and T.~Fülöp.
\newblock Implicit numerical schemes for generalized heat conduction equations.
\newblock {\em International Journal of Heat and Mass Transfer}, 126:1177 --
  1182, 2018.
  
\bibitem{IgnOst10b}
J.~Ignaczak, M.~Ostoja-Starzewski.
\newblock Thermoelasticity with finite wave speeds.
\newblock {\em Oxford University Press, Oxford}, 2010.

\bibitem{SellCimm19a}
A.~Sellitto, V. A.~Cimmelli.
\newblock Heat-pulse propagation in thermoelastic systems: application to graphene.
\newblock {\em Acta Mechanica}, 230(1):121--136, 2019.

\bibitem{SellCimm19b}
A.~Sellitto, V. A.~Cimmelli, D.~Jou.
\newblock Nonlinear Propagation of Coupled First-and Second-Sound Waves in Thermoelastic Solids.
\newblock {\em Journal of Elasticity}, 1--17, 2019

\bibitem{BarZan97}
A.~Barletta, E.~Zanchini.
\newblock Hyperbolic heat conduction and local equilibrium: a second law analysis.
\newblock {\em International Journal of Heat and Mass Transfer}, 40(5):1007 --
  1016, 1997.

\bibitem{BarZan98}
A.~Barletta, E.~Zanchini.
\newblock Nonequilibrium temperature and hyperbolic heat conduction.
\newblock {\em Physical Review B}, 57(22):14228, 1998.

\bibitem{Zan99}
E.~Zanchini.
\newblock Hyperbolic-heat-conduction theories and nondecreasing entropy.
\newblock {\em Physical Review B}, 60(2):991, 1999.

\bibitem{Romano}
G. Mascali, V. Romano.
\newblock Charge Transport in Graphene including Thermal Effects.
\newblock {\em SIAM Journal on Applied Mathematics}, 77(2), 593-613, 2017.


\bibitem{ColNew88nl}
B.~D. Coleman and D.~C. Newman.
\newblock Implications of a nonlinearity in the theory of second sound in
  solids.
\newblock {\em Physical Review B}, 37(4):1492, 1988.



\bibitem{Jordan15}
P.~M. Jordan.
\newblock Second-sound propagation in rigid, nonlinear conductors.
\newblock {\em Mechanics Research Communications}, 68:52--59, 2015.

\bibitem{Kirc1894b}
G.~Kirchhoff.
\newblock {\em Vorlesungen {\"u}ber die {T}heorie der {W}{\"a}rme}, volume~4.
\newblock BG Teubner, 1894.



\bibitem{Grof18}
Gy. Gr{\'o}f.
\newblock Notes on using temperature-dependent thermal diffusivity - forgotten
  rules.
\newblock {\em Journal of Thermal Analysis and Calorimetry}, 132(2):1389--1397,
  2018.




\bibitem{Ons}
L. Onsager.
\newblock Reciprocal Relations in {I}rreversible {P}rocesses.
\newblock {\em Phys. Rev.},37,  419, 1931.


\bibitem{CimJouRugVan}
V. A. Cimmelli, D. Jou,  T. Ruggeri and P. V\'an.
\newblock{Entropy {P}rinciple and {R}ecent {R}esults in {N}on-{E}quilibrium {T}heories}.
\newblock{\em Entropy}, 16, 1758-1807, 2014.
 
 
 
\bibitem{BCTFGGAGPV13}
B.~Cz\'el, T.~Fülöp, Gy. Gr\'of, \'A. Gyenis, and P.~V\'an.
\newblock Simple heat conduction experiments.
\newblock In Dombi Sz., editor, {\em 11th International Conference on Heat
  Engines and Environmental Protection}, pages 141--146, Budapest, 2013. BME,
  Dep. of Energy Engineering.

\bibitem{ParEtal61}
W.~J. Parker, R.~J. Jenkins, C.~P. Butler, and G.~L. Abbott.
\newblock Flash method of determining thermal diffusivity, heat capacity, and
  thermal conductivity.
\newblock {\em Journal of Applied Physics}, 32(9):1679--1684, 1961.

\bibitem{James80}
H.~M. James.
\newblock Some extensions of the flash method of measuring thermal diffusivity.
\newblock {\em Journal of Applied Physics}, 51(9):4666--4672, 1980.

\bibitem{Kov18gk}
R.~Kovács.
\newblock Analytic solution of {G}uyer-{K}rumhansl equation for laser flash
  experiments.
\newblock {\em International Journal of Heat and Mass Transfer}, 127:631--636,
  2018.

\bibitem{WeickEtal98}
J.~Weickert, Bart M.~H. Romeny, and M.~A. Viergever.
\newblock Efficient and reliable schemes for nonlinear diffusion filtering.
\newblock {\em IEEE Transactions on Image Processing}, 7(3):398--410, 1998.

\bibitem{Weick96a}
J.~Weickert.
\newblock Nonlinear diffusion scale-spaces: {F}rom the continuous to the
  discrete setting.
\newblock In {\em ICAOS'96}, pages 111--118. Springer, 1996.

\bibitem{Weick97rev}
J.~Weickert.
\newblock A review of nonlinear diffusion filtering.
\newblock pages 1--28, 1997.

\bibitem{NumRec07b}
W.~H. Press.
\newblock {\em Numerical {R}ecipes 3rd {E}dition: {T}he {A}rt of {S}cientific
  {C}omputing}.
\newblock Cambridge University Press, 2007.

\bibitem{Jury74}
E.~I. Jury.
\newblock Inners and {S}tability of {D}ynamic systems. 1974.

\end{thebibliography}



\end{document}